\documentstyle[prd,aps]{revtex}

\input epsf

\newcommand{\bea}{\begin{eqnarray}}
\newcommand{\eea}{\end{eqnarray}}

\begin{document}
\twocolumn[\hsize\textwidth\columnwidth\hsize\csname
@twocolumnfalse\endcsname

\title{Non-singular big-bounces and evolution of linear fluctuations}
\author{Jai-chan Hwang${}^{(a,b)}$ and Hyerim Noh${}^{(c,b)}$ \\
        ${}^{(a)}$ {\sl Department of Astronomy and Atmospheric Sciences,
                        Kyungpook National University, Taegu, Korea} \\
        ${}^{(b)}$ {\sl Institute of Astronomy, Madingley Road, 
                        Cambridge, UK} \\
        ${}^{(c)}$ {\sl Korea Astronomy Observatory, Taejon, Korea} 
        }

\date{\today}
\maketitle

\begin{abstract}

We consider evolutions of {\it linear} fluctuations as the background
Friedmann world model goes from contracting to expanding phases through
smooth and non-singular bouncing phases.
As long as the gravity dominates over the pressure gradient 
in the perturbation equation the growing-mode in the expanding phase 
is characterized by a conserved amplitude, we call it a $C$-mode. 
In the spherical geometry with a pressureless medium, we show that
there exists a special gauge-invariant combination $\Phi$ which stays
constant throughout the evolution from the big-bang to the big-crunch
with the same value even after the bounce:
it characterizes the coefficient of the $C$-mode.
We show this result by using a bounce model where the pressure gradient 
term is negligible during the bounce; this requires additional presence 
of an exotic matter.
In such a bounce, even in more general situations of the equation of states
before and after the bounce, the $C$-mode in the expanding phase is affected 
only by the $C$-mode in the contracting phase, thus the growing mode in 
the contracting phase decays away as the world model enters expanding phase.
In the case the background curvature has significant role during
the bounce, the pressure gradient term becomes important and we cannot 
trace $C$-mode in the expanding phase to the one before the bounce.
In such situations, perturbations in a fluid bounce model show 
exponential instability, whereas the ones in a scalar field bounce model 
show oscillatory behaviors.

\end{abstract}


\vskip2pc]

\section{Introduction}
                                                  \label{sec:Introduction}

The collapsing and bouncing phases of the 
FLRW (Friedmann-Lema\^itre-Robertson-Walker) world models
we are considering are the possible ones in our past, before the big-bang.
The same physics, however, could work in the possible case in future as well.

The re-expansion of a positive curvature Friedmann world model
which is destined to collapse, or cyclic repetition of the process
with such a bounce was proposed as early as in the 1930's 
\cite{Tolman-1931,Tolman-1934,Peebles-1993}.
Specific realizations of the bounce and the conditions required 
to have the FLRW world model from a bounce were studied in 
\cite{Nariai-1971,Molina-Paris-Visser-1999}.
The singularity-free cosmologies are possible as we give up the 
strong energy condition which is often possible with quantum corrections
\cite{Nariai-1971}.
Recently, the big-bang world model preceded by a collapsing
phase attracted renewed attention in the context of the
brane cosmology \cite{bounce-cyclic}.

In this paper we analyse the evolution of scalar-type 
curvature (often called the adiabatic) fluctuations
as the background world model goes through a smooth and non-singular
bounce which connects the contracting and the expanding phases.
We will {\it assume} the classical General Relativity is valid
as the correct gravity theory throughout the evolution,
and also consider scales where the linear approximation is valid.

In \S \ref{sec:Review} we review the cosmological perturbation theory
needed for our analyses in later sections.
In \S \ref{sec:Solutions} we present the large-scale evolutions
of various curvature perturbations near singularity.
In \S \ref{sec:Pressureless} we analyse the evolution using exact
solutions in a pressureless situation.
In \S \ref{sec:Bounce} we show the evolution of perturbation through
a bounce model using three different bounce models.
\S \ref{sec:Discussions} presents a summary, implications of our work,
and discussions on related works.
We set $c \equiv 1$.

\section{Cosmological perturbations}
                                              \label{sec:Review}

Our metric convention is \cite{Bardeen-1988,HN-2002-CMBR}
\bea
   ds^2
   &=& - a^2 (1 + 2 \alpha) d\eta^2
       - 2 a^2 ( \beta_{,\alpha} + B_\alpha^{(v)} ) d \eta d x^\alpha
   \nonumber \\
   & & + a^2 \Big[ g^{(3)}_{\alpha\beta} ( 1 + 2 \varphi )
       + 2 \gamma_{,\alpha|\beta} + 2 C^{(v)}_{(\alpha|\beta)}
   \nonumber \\
   & & \qquad
       + 2 C^{(t)}_{\alpha\beta} \Big] dx^\alpha dx^\beta.
\eea
The perturbed order variables $\alpha$, $\beta$, $\varphi$ and
$\gamma$ are scalar-type perturbations;
the transverse $B^{(v)}_\alpha$ and $C^{(v)}_\alpha$ are
vector-type perturbations (rotation);
a transverse-tracefree $C^{(t)}_{\alpha\beta}$ is tensor-type
perturbation (gravitational wave).
The energy-momentum tensor is
\bea
   T^0_0
   &\equiv& - (\bar \mu + \delta \mu),
   \nonumber \\
   T^0_\alpha
   &\equiv& (\mu + p) \left[ - (1/k) v_{,\alpha} + v^{(v)}_\alpha \right],
   \nonumber \\
   T^\alpha_\beta
   &\equiv& (\bar p + \delta p) \delta^\alpha_\beta + \pi^\alpha_\beta,
\eea
where the tracefree $\pi^\alpha_\beta$ is the anisotropic stress.

The trace and tracefree parts of extrinsic curvature
(equivalently, the expansion $\hat \theta$ and the shear $\hat
\sigma_{ab}$ of the normal frame vector field), and the intrinsic
scalar curvature $R^{(h)}$ of the constant-time spacelike hypersurface
are, see eqs. (C3,C14) of \cite{HN-2002-CMBR} and eqs. (A6,A7) of
\cite{Bardeen-1980},
\bea
   \hat \theta
   &=& 3 H - \kappa,
   \nonumber \\
   \hat \sigma_{\alpha\beta}
   &=& \chi_{,\alpha|\beta}
       - {1 \over 3} g^{(3)}_{\alpha\beta} \Delta \chi
       + a \Psi^{(v)}_{(\alpha|\beta)}
       + a^2 \dot C^{(t)}_{\alpha\beta},
   \nonumber \\
   R^{(h)}
   &=& {1 \over a^2} \left[ 6 K - 4 (\Delta + 3 K) \varphi \right],
\eea
where\footnote{
     Indices of $\hat \sigma_{\alpha\beta}$ 
     ($E_{\alpha\beta}$ and $H_{\alpha\beta}$, later) 
     are based on the spacetime metric $g_{ab}$.
     All the other Greek indices are based on the
     $g^{(3)}_{\alpha\beta}$.
     }
\bea
   & & \chi \equiv a (\beta + a \dot \gamma ), \quad
       \kappa \equiv 3 ( H \alpha - \dot \varphi )
       - {\Delta \over a^2} \chi,
   \nonumber \\
   & & \Psi^{(v)}_\alpha \equiv B^{(v)}_\alpha + a \dot C^{(v)}_\alpha.
\eea
We have $H \equiv {\dot a \over a}$, $K$ the normalized background
three-space curvature, and an overdot indicates time derivative
based on $t$, $dt \equiv a d \eta$. Thus, $\kappa$, $\chi$, and
$\varphi$ are the perturbed expansion, the scalar-type shear, and
the perturbed three-space scalar-curvature of the normal
hypersurface, respectively.
$\Psi^{(v)}_\alpha$ and $\dot C^{(t)}_{\alpha\beta}$ give the
vector- and tensor-type contributions to the shear tensor.

$C^{(t)}_{\alpha\beta}$, $\Psi^{(v)}_\alpha$ and $v^{(v)}_\alpha$
are gauge-invariant.  $\alpha$, $\varphi$, $\chi$, $\kappa$, $v$,
$\delta \mu$ and $\delta p$ are spatially gauge-invariant but
depend on the temporal gauge condition, i.e, depend on the spatial
hypersurface (time slicing) choice \cite{Bardeen-1980}. Setting
any one of these temporally gauge dependent variables equal to zero
corresponds to a fundamental gauge condition; except for the
synchronous gauge ($\alpha \equiv 0$) each of the other conditions
fixes the temporal gauge degree of freedom completely, and any
variable in such a gauge condition uniquely corresponds to a
gauge-invariant combination (of the variable and the variable used
in the gauge condition) \cite{Bardeen-1988,HN-2002-CMBR}.

Equations describing the evolution of a spatially homogeneous and
isotropic FLRW world model are:
\bea
   & & H^2 = {8 \pi G \over 3} \mu - {K \over a^2} + {\Lambda \over 3}, \quad
       \dot \mu = - 3 H ( \mu + p).
   \label{BG-eqs}
\eea
The $\Lambda$ term can be also considered as an ideal fluid with
$\mu_\Lambda = - p_\Lambda = \Lambda/(8 \pi G)$;
in such cases, we have $\mu = \mu_m + \mu_\Lambda$ and $p = p_m + p_\Lambda$.
In the case of a minimally coupled scalar field ($\phi$) we have
$\mu = {1 \over 2} \dot \phi^2 + V$
and $p = {1 \over 2} \dot \phi^2 - V$, with an equation of motion 
$\ddot \phi + 3 H \dot \phi + V_{,\phi} = 0$.
$\Lambda$ can be included as $V_\Lambda = \Lambda/(8 \pi G)$.

We consider the general scalar-type perturbation.
It is convenient to introduce 
\cite{Field-Shepley-1968,Hwang-Vishniac-1990,Hwang-1999}
\bea
   & & \Phi \equiv \varphi_v - {K /a^2 \over 4 \pi G ( \mu + p) } \varphi_\chi.
   \label{Phi}
\eea
The scalar-type perturbation of a fluid with vanishing anisotropic stress
in the Einstein's gravity is described by \cite{Hwang-1999}:
\bea
   & & \Phi = {H^2 \over 4 \pi G ( \mu + p) a}
       \left( {a \over H} \varphi_\chi \right)^\cdot, 
   \label{Phi-varphi_chi-1} \\
   & & \dot \Phi = - {H c_s^2 \over 4 \pi G ( \mu + p) } {k^2 \over a^2}
       \varphi_\chi - {H \over \mu + p} e, 
   \label{Phi-varphi_chi-2} \\
   & & {k^2 - 3 K \over a^2} \varphi_\chi = 4 \pi G \mu \delta_v,
   \label{Poisson-eq}
\eea
where $w \equiv p/\mu$ and $c_s^2 \equiv \dot p/\dot \mu$.
$\varphi_v \equiv \varphi -
(aH/k) v$, $\varphi_\chi \equiv \varphi - H \chi$ and $\delta_v
\equiv \delta + 3 (aH/k) (1 + w) v$ are gauge-invariant
combinations \cite{Bardeen-1988,HN-2002-CMBR};
$\varphi_v$ is the same as $\varphi$ in the comoving
gauge ($v \equiv 0$), and $\varphi_\chi$ is the same as $\varphi$
in the zero-shear gauge ($\chi \equiv 0$), etc.

We emphasize that results in this section are valid considering general
$K$, $\Lambda$, and time-varying equation of state $p(\mu)$.
In the case of a minimally coupled scalar field, we have additional
nonvanishing entropic perturbation (the isotropic stress), 
$e = (1 - c_s^2) \delta \mu_v$.
Its effect can be covered by changing $c_s^2$ in 
eq. (\ref{Phi-varphi_chi-2}) to 
$c_A^2 \equiv 1 - 3 ( 1- c_s^2) K/k^2$, 
\cite{Hwang-Noh-2001-Fluids};
as eqs. (\ref{Phi-varphi_chi-1}-\ref{Poisson-eq}) are a complete set for 
single component, this prescription applies always
in the single scalar field case.
It is convenient to have 
\bea
   & & \varphi_\delta \equiv \varphi + {\delta \over 3 (1 + w)}
       = \Phi + {k^2 \over a^2}
       {1 \over 12 \pi G ( \mu + p)} \varphi_\chi.
   \label{varphi_delta}
\eea
We can show in general \cite{Hwang-Noh-2001-Fluids}
\bea
   & & \varphi_\kappa = \varphi_\delta \Big/ \left[ 1 
       + {k^2 - 3K \over 12 \pi G ( \mu + p) a^2} \right],
   \label{varphi_kappa}
\eea
where $\varphi_\kappa \equiv \varphi + H \kappa/(3 \dot H - k^2/a^2)$.
In the notation of Bardeen \cite{Bardeen-1980} we have
\bea
   & & \delta_v = \epsilon_m, \quad
       \varphi_\chi = \Phi_H, \quad
       \varphi_v = \phi_m, \quad
       \varphi_\kappa = \phi_h. 
\eea       
We have $\varphi_\delta = \zeta$ in \cite{Bardeen-1988}, 
and $\varphi_v = {\cal R}$ in \cite{Liddle-Lyth-2000};
$\varphi_v$ was also originally introduced by Lukash 
as $-{1 \over 3} q$ in \cite{Lukash-1980}.
{}From eqs. (\ref{Phi-varphi_chi-1},\ref{Phi-varphi_chi-2}) 
we can derive equations in closed forms:
\bea
   & & \ddot {\bar \Phi} + \left( c_s^2 {k^2 / a^2} - \ddot x / x \right) 
       \bar \Phi = 0, 
   \label{Phi-eq} \\
   & & \ddot {\bar \varphi}_\chi + \left( c_s^2 {k^2 / a^2}
       - \ddot y / y \right) \bar \varphi_\chi = 0, 
   \label{varphi_chi-eq} 
\eea
where 
\bea
   & & \bar \Phi \equiv x \Phi, \quad
       \bar \varphi_\chi \equiv ({a y / H}) \varphi_\chi, 
   \nonumber \\
   & & y \equiv H/\sqrt{( \mu + p) a}
       \equiv (a / c_s) x^{-1}.
\eea
Equations using the conformal time were presented in
\cite{Mukhanov-etal}.

In the large-scale limit (meaning $c_s^2 k^2/a^2$ term negligible,
thus gravity dominates over pressure) we have general solutions
\cite{Hwang-Vishniac-1990,Hwang-1999,Hwang-Noh-2001-Fluids}:
\bea
   & & \Phi (k,t) = C (k) 
       - d (k) {k^2 \over 4 \pi G} \int^t dt/ x^2,
   \label{Phi-sol} \\
   & & \varphi_\chi (k,t) 
       = 4 \pi G C (k) {H \over a} \int^t dt/y^2 + d (k) {H \over a}, 
   \label{varphi_chi-sol} 
\eea
where $C$ and $d$ are the two spatially dependent integration constants:
we call these the $C$-mode and the $d$-mode, respectively.
Solutions for $\delta_v$, $\varphi_\delta$ and $\varphi_\kappa$
follow from eqs. (\ref{Poisson-eq}-\ref{varphi_kappa}).
Notice that the $d$-mode of $\Phi$
is higher-order in the large-scale expansion
compared with the $d$-mode of $\varphi_\chi$.
In a pressureless medium, the above
solutions are exact, and we have $\Phi = C$ \cite{Field-Shepley-1968}.
In fact, for such a medium, instead of eq. (\ref{Phi-eq}),
eq. (\ref{Phi-varphi_chi-2}) gives $\dot \Phi = 0$.

In order to use the large-scale solutions in 
eqs. (\ref{Phi-sol},\ref{varphi_chi-sol}) it is important to check 
whether we could ignore the $c_s^2 k^2/a^2$ term during the evolution.
The large-scale condition implies 
$({{\rm pressure} / {\rm gravity}}) \ll 1$ where
\bea
   & & {{\rm pressure} \over {\rm gravity}}
       \quad \sim \quad {c_s^2 k^2/a^2 \over \ddot x/x}, \quad
       {c_s^2 k^2/a^2 \over \ddot y/y}.
   \label{pressure}
\eea
In a positive curvature (spherical) model the wave number varies
as $k = \sqrt{(n^2 - 1)K}$ where $n = 1, 2, 3, \dots$;
$n = 1, 2$ are known to be unphysical \cite{Lifshitz-1946,Bardeen-1980}. 
In a negative curvature (hyperbolic) model, $k > \sqrt{|K|}$ and
$0 \le k < \sqrt{|K|}$ correspond to the subcurvature and the supercurvature
scales, respectively \cite{Lyth-Woszczyna-1995}.
In a zero curvature (flat) model we have $k \ge 0$.

The following two variables are continuous under a sudden jump
of equation of state \cite{Hwang-Vishniac-1991}:
\bea
   & & \varphi_\chi \quad ({\rm or} \; \delta_v), \quad
       \varphi_\delta.
   \label{matching-variables}
\eea
These joining variables work for general $K$, $\Lambda$,
and $p(\mu)$ in the general scale.
This applies for the perfect fluids, and for the cases involving 
scalar fields, see \cite{Hwang-Vishniac-1991}.
{}For the background, $a$ and $\dot a$ should be continuous at the transition.
Consider two phases $I$ and $II$ with different equation of states, 
$w_I$ and $w_{II}$, making a transition at $t_1$.
Assuming a flat background, in the large-scale limit, 
by matching $\varphi_\chi$ and $\varphi_\delta$ in 
eqs. (\ref{Phi-sol},\ref{varphi_chi-sol},\ref{varphi_delta}) 
we can see that to the leading order in the large-scale expansion we have
\bea
   & & C_{II} = C_{I},
   \nonumber \\
   & & d_{II} = d_{I} + 4 \pi G C_{I} \Bigg[ 
       \int^{t_1} {a (\mu + p) \over H^2} dt \Bigg|_{I}
   \nonumber \\
   & & \qquad
       - \int^{t_1} {a (\mu + p) \over H^2} dt \Bigg|_{II}
       \Bigg].
   \label{C-1}
\eea
Thus, to the leading order in the large-scale expansion
the $C$-mode of $\Phi$ remains the same, whereas
the $d$-mode of $\varphi_\chi$ is affected by the transition
and also the previous history of the $d$- and $C$-modes
\cite{Hwang-Vishniac-1991}.
Applications were made in \cite{joining-applications}.

Ignoring the anisotropic stress ($\pi^\alpha_\beta$) and assuming $K = 0$, 
equations for the rotation and the gravitational wave become
\cite{HN-2002-CMBR}
\bea
   & & \big[ a^4 (\mu + p) v^{(v)}_\alpha \big]^\cdot = 0; \quad
       {k^2 \over a^2} \Psi^{(v)}_\alpha = 16 \pi G ( \mu + p) v^{(v)}_\alpha,
   \label{rotation-eq} \\
   & & \ddot {\bar C}^{(t)}_{\alpha\beta} + \left( k^2 / a^2
       - \ddot z / z \right) \bar C^{(t)}_{\alpha\beta} = 0,
   \label{GW-eq}
\eea
where $\bar C^{(t)}_{\alpha\beta} \equiv z C^{(t)}_{\alpha\beta}$ and
$z \equiv a^{3/2}$.
The equation for $C^{(t)}_{\alpha\beta}$ satisfies the same
equation as $\Phi$ in eq. (\ref{Phi-eq}) with $x \propto a^{3/2}$.
Thus, in general scale for $\Psi^{(v)}_\alpha$
and in the large-scale limit for $C^{(t)}_{\alpha\beta}$ we have
general solutions
\bea
   & & \Psi^{(v)}_\alpha (k, t) = d_\alpha (k) {1 \over a^2},
   \label{rotation-sol} \\
   & & C^{(t)}_{\alpha\beta} (k,t) = c_{\alpha\beta} (k) 
       - d_{\alpha\beta} (k) \int^t dt/a^3.
   \label{GW-sol}
\eea
Similarly as the $C$-mode of $\Phi$
the amplitude of $c_{\alpha\beta}$-mode simply stays constant.

\section{Large-scale evolutions of curvature perturbations}
                                              \label{sec:Solutions}

The general large-scale solutions for scalar- and tensor-type perturbations, 
and the general solutions for the vector-type perturbation
are presented in eqs. 
(\ref{Phi-sol},\ref{varphi_chi-sol},\ref{rotation-sol},\ref{GW-sol}).
We call the solutions with $C$ and $c_{\alpha\beta}$ the $C$-modes,
and the solutions with $d$, $d_\alpha$ and
$d_{\alpha\beta}$ the $d$-modes.  The vector-type perturbation
has no $C$-mode.  
In the expanding phase $C$-modes are relatively growing solutions 
whereas $d$-modes decay, thus transient, in time. 
In a contracting phase, however, the opposite is the case
with the $d$-modes often diverging as the background model
approaches the singularity.

In this section we assume near {\it flat} background.
With a constant $w$ we have $c_s^2 = w$, and 
$a \propto |t|^{2 \over 3(1+w)}$ for $-1<w \le 1$.  
In a medium with $w>- {1\over 3}$, as we approach the singularity
${k \over aH} ( \propto |t|^{1+3w \over 3 + 3w})$ becomes negligible
for any given scale with $k$, thus the large-scale conditions
are well satisfied.
In such a case, during dynamical time-scale of the background evolution
``light can travel only a small fraction of a wavelength''
\cite{Bardeen-1980}, thus the scale becomes super-horizon scale;
Bardeen \cite{Bardeen-1980} called it (${k \over aH} \sim 1$)
an ``effective particle horizon''\footnote{
      The effective particle horizon is the same as the
      ``Hubble sphere'' studied
      in \cite{Harrison-1991}, and closely resembles the ``$z$-surface''
      introduced in \cite{Lightman-Press-1989}.
      The global concepts like particle- and event-horizons
      are not suitable to describe the local dynamically reachable ranges.
      Studies in \cite{Harrison-1991,Lightman-Press-1989} show that
      the Hubble sphere and the $z$-surface are more suitable to
      describe the concepts like `scales becomming super-horizon
      size during the inflation era'.
      Similarly, these are suitable to describe the same physics
      during contracting phase with $w > - {1 \over 3}$.
      Under this situation we can show that to an observer
      in the contracting phase,
      an object separated by a given comoving distance appears
      more blue-shifted as time goes on, i.e, the object is
      effectively receding from the observer.
      }.
In general, we consider $w \ge 0$: ``a single-component treatment of
the matter is inappropriate when the net pressure is negative''
\cite{Bardeen-1980}.

In this case, from eqs.
(\ref{Phi-sol},\ref{varphi_chi-sol},\ref{GW-sol}) we have the
$C$-modes remain constant in time
\bea
   & & \varphi_\chi = {3 + 3 w \over 5 + 3w} C, \quad
       \varphi_v = C, \quad
       C^{(t)}_{\alpha\beta} = c_{\alpha\beta}.
   \label{C-mode-sol1}
\eea
Thus, in an expanding medium the perturbation evolutions in the
super-effective-particle-horizon are 
kinematic\footnote{
     This kinematic nature of the evolution is reflected
     in the conserved behaviors of linear perturbations in the
     expanding phase \cite{Bardeen-1980,Bardeen-1988}.
     What is conserved is the the amplitude of the $C$-mode
     of curvature perturbation in diverse gauge conditions.
     In the expanding phase the situation is also well
     represented by the
     separated Friedmann picture of the perturbed Friedmann
     world model pioneered (up to our knowledge)
     by Harrison in \cite{Harrison-1970}.
     }
and are characterized by the conserved quantities, see
\cite{Starobinsky-etal-2001}.

Meanwhile, the $d$-modes behave as
\bea
   & & \varphi_\chi \propto |t|^{-{5+3w \over 3+3w}}, \quad
       \varphi_v, \; C^{(t)}_{\alpha\beta}
       \propto |t|^{-{1-w \over 1+w}},
   \nonumber \\
   & & \Psi^{(v)}_\alpha \propto |t|^{-{4 \over 3 (1 + w)}},
\eea
for $-1< w < 1$. 
{}For $w=1$, the above solution is valid for
$\varphi_\chi (\propto |t|^{-4/3}$) and $\Psi^{(v)}_\alpha
(\propto |t|^{-2/3}$), whereas we have $\varphi_v, \;
C^{(t)}_{\alpha\beta} \propto \ln{|t|}$ instead.

In the case of constant $w$, a complete set of solutions of the
scalar-type perturbation in six different fundamental gauge
conditions is presented in Tables 2-5 of \cite{Hwang-1993-IF}.
Although the solutions in \cite{Hwang-1993-IF} are presented in
the context of the expanding phase, by changing the time variables
to their absolute values with the singularity at $|t| = 0$, the
same solutions apply in the contracting phase as well.

\subsection{Intrinsic curvature}

$\varphi$ is a dimensionless measure of the (intrinsic) curvature perturbation
of the hypersurface (temporal gauge condition) we choose. Thus,
its value depends on the chosen hypersurface (temporal gauge condition). 
{}From eq. (A6) of \cite{Bardeen-1980} we
notice that $C^{(t)}_{\alpha\beta}$ gives a dimensionless
contribution to the tensor-type intrinsic curvature perturbation,
and the vector-type perturbation does not contribute to the
curvature perturbation, see also eq. (C14) in \cite{HN-2002-CMBR}.
Table 2 of \cite{Hwang-1993-IF} shows that for the
$d$-mode\footnote{
     Although $\varphi_\alpha$, the $\varphi$ in the synchronous gauge
     ($\alpha \equiv 0$), is not gauge-invariant, we are considering the
     physical solutions.}
\bea
   & & \varphi_\kappa, \; \varphi_\alpha, \; \varphi_\delta, \; \varphi_v, \;
       C^{(t)}_{\alpha\beta} \propto |t|^{-{1-w \over 1+w}}, \quad
       \varphi_\chi \propto |t|^{-{5 + 3w \over 3 + 3w}},
   \label{curvature-sol}
\eea
for $-1< w <1$.
{}For $w = 1$ we have\footnote{
     One other special case occurs for
     $w = 0$ where we have no $d$-modes for $\varphi_v$ and
     $\varphi_\alpha$, see Table 5 of \cite{Hwang-1993-IF}.
     },
from Table 4 of \cite{Hwang-1993-IF},
\bea
   & & \varphi_\kappa, \; \varphi_\alpha, \; \varphi_\delta, \;
       \varphi_v, \; C^{(t)}_{\alpha\beta} \propto \ln{|t|}, \quad
       \varphi_\chi \propto |t|^{-4/3}.
   \label{curvature-sol-2}
\eea
Thus, even for $w = 1$, $\varphi$ diverges in all gauge conditions
considered. $\varphi$ in the zero-shear gauge diverges more
strongly compared with the ones in the other gauge conditions. The
strong divergence in the zero-shear gauge is known to be due to
the strong curvature of the hypersurface (temporal gauge
condition) \cite{Bardeen-1980}.  $\varphi$ is set to zero in the
uniform-curvature gauge.  {}For the $C$-modes we have eq.
(\ref{C-mode-sol1}) and $\varphi_\kappa = \varphi_\alpha =
\varphi_\delta = \varphi_v = C$.

\subsection{Extrinsic curvature}

Although we have no shear in the background model, the perturbed
scalar-type shear is still a gauge dependent quantity.
A dimensionless measure of the shear variable
(the shear divided by the background expansion rate) becomes
\bea
   {\hat \sigma \over H}
   &\sim& \left( {k^2 \over a^2 H^2} H \chi, \;
       {k \over aH} \Psi^{(v)}_\alpha, \;
       {1 \over H} \dot C^{(t)}_{\alpha\beta} \right),
\eea
for three perturbation types;
$\hat \sigma \equiv
 \sqrt{ \hat \sigma_{\alpha\beta} \hat \sigma^{\alpha\beta} /2 }$.
The $d$-modes of all-types of
perturbations show {\it the same} temporal behavior
\bea
   {\hat \sigma \over H} \propto |t|^{-{1-w \over 1+w}},
\eea
for $-1<w \le 1$, thus, with no logarithmic divergences for $w=1$
case.  The result for vector- and tensor-type perturbations follow
from eqs. (\ref{rotation-sol},\ref{GW-sol}), and the one
for scalar-type perturbation follows from the Tables 2 and 4 of
\cite{Hwang-1993-IF}; thus, the solutions apply to the gauge
conditions considered (we set $\chi = 0$ in the zero-shear gauge).
We can check that the $C$-modes contributions to $\hat \sigma/H$
are all regular near the singularity for $-1<w \le 1$. The
behavior of $\kappa/H$, a dimensionless measure of the perturbed
trace part of the extrinsic curvature, varies widely depending on
the gauge conditions, see Tables 2 and 4 of \cite{Hwang-1993-IF}.

\subsection{Weyl curvature}

Durrer has informed us another useful measure of
the spacetime fluctuation which behaves regularly for $w=1$,
the Weyl curvature $C_{abcd}$.
The Weyl-curvature (the conformal tensor) vanishes in the
FLRW background geometry, and is naturally gauge-invariant.
The Weyl tensor can be covariantly decomposed to the
electric $E_{ab}$ and magnetic $H_{ab}$ parts, \cite{covariant}.
Using eq. (C9) in \cite{HN-2002-CMBR},
see also eqs. (2.26,2.27) of \cite{Durrer-1994}, we can show
\bea
   {E \over R}
   &\sim& \left( {k^2 \over a^2 H^2} \varphi_\chi, \;
       {k \over aH^2} \dot \Psi^{(v)}_\alpha, \;
       {1 \over H} \dot C^{(t)}_{\alpha\beta} \right),
\eea
where $E \equiv \sqrt{E^{\alpha\beta} E_{\alpha\beta}/2}$ and
$R$ is the scalar-curvature ($\sim H^2$).
Thus, the $d$-modes of all-types of perturbations behave
exactly like $\hat \sigma/H (\propto |t|^{-{1-w \over 1+w}})$,
thus behave regularly for $w = 1$.
$H_{ab}$ only contributes to the vector- and tensor-type perturbations,
and we can also show that the $d$-modes
behave as $\sqrt{H^{\alpha\beta} H_{\alpha\beta}/2}/R \propto
 |t|^{- {2 \over 3} {1 - 3w \over 1 + w}}$, thus
behave more regularly for $w = 1$.
This regular behavior of the Weyl curvature at singularity for
$w=1$ case was used to argue the validity of
perturbation theory in such a situation,
see around eq. (5.20) of \cite{Brustein-etal-1995};
however, see our discussion below eq. (\ref{Lyth-conditions}).

\section{Exact evolution in a pressureless case}
                                              \label{sec:Pressureless}

{}For $K > 0$, $\Lambda = 0$ and $p = 0$, eq. (\ref{BG-eqs}) gives 
a cycloid \cite{Friedmann-1922,Tolman-1934}
\bea
   & & a = {c_m } \left( 1 - \cos{\eta} \right), \quad
       t = {c_m } \left( \eta - {\sin{\eta} } \right),
   \label{BG-sol}
\eea
where $c_m \equiv (4 \pi G/3) \mu a^{3}$ and $d \eta \equiv dt/a$.
$K$ is normalized to unity, thus $0 \le \eta \le 2 \pi$.

{}For $p = 0$, eqs. (\ref{Poisson-eq},\ref{varphi_chi-eq}) give
\bea
   & & \left[ a^2 H^2 \left( {\delta_v / H} \right)^\cdot \right]^\cdot 
       / (a^2 H)
   \nonumber \\
   & & = \ddot \delta_v + 2 H \dot \delta_v + 4 \pi G \mu \delta_v = 0,
   \label{delta_v-eq-MDE}
\eea
which coincides with the density perturbation equation in the synchronous 
gauge \cite{Lifshitz-1946}, or in the Newtonian context \cite{Jeans-1902};
$\delta_v$ is the energy density perturbation in the comoving gauge,
\cite{Nariai-1969}.
Assuming $\Lambda = 0$, the two independent {\it exact} solutions 
for $\varphi_\chi \propto \delta_v/a$ in eq. (\ref{varphi_chi-sol}) are 
\cite{Weinberg-1972}:
\bea
   & & c_m {H \over a} \int^t {dt \over a^2 H^2}
       = - {3 \eta \sin{\eta} \over ( 1 - \cos{\eta} )^3} 
       + {5 + \cos{\eta} \over (1 - \cos{\eta})^2} 
   \nonumber \\
   & & \qquad
       \equiv \varphi_+ (\eta)/3,
   \nonumber \\
   & & c_m^2 {H \over a} 
       = {\sin{\eta} \over ( 1 - \cos{\eta} )^3} \equiv \varphi_d (\eta).
\eea
In asymptotes we have
\bea
   & & \varphi_+ (\eta) \simeq {3 / 5}, \quad
       \varphi_d (\eta) \simeq {8 / \eta^5}, \quad 
       (\eta \ll 1),
   \nonumber \\
   & & \varphi_+ (\eta) / (18 \pi) \simeq (8 / \bar \eta^5) 
       \simeq - \varphi_d (\eta), \quad
       (\bar \eta \ll 1), \quad
   \label{asymptotes}
\eea
where $\bar \eta \equiv 2 \pi - \eta$.
We have
\bea
   & & \varphi_+ (2 \pi - \eta) = \varphi_+ (\eta) + 18 \pi \varphi_d (\eta)
       \equiv \varphi_- (\eta),
   \label{varphi_-}
\eea
where $\varphi_-$ shows time inverted evolution of $\varphi_+$,
\cite{Groth-Peebles-1975}.

\begin{figure}[ht]
   \centering
   \leavevmode
   \epsfysize=8cm
   \epsfbox{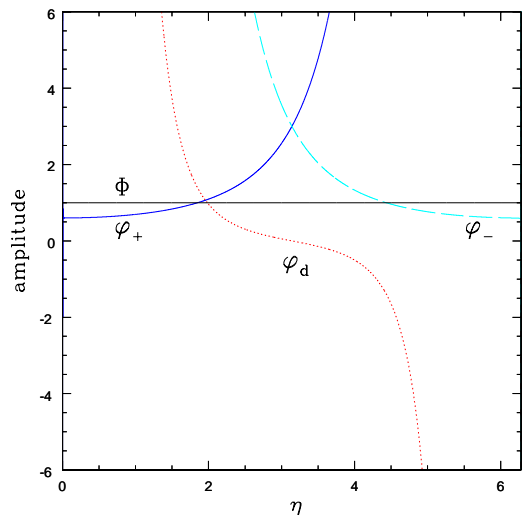}\\
   \caption[fig]
   {\label{Fig-pressureless-pert}
   Right: Evolutions of $\varphi_{+}$ (blue, line), 
          $\varphi_{-}$ (cyan, dashed line), 
          $\varphi_{d}$ (blue, dotted line), and 
          $\Phi$ (black, horizontal line).
    }
\end{figure}

Equations 
(\ref{Phi-sol},\ref{varphi_chi-sol},\ref{Poisson-eq}-\ref{varphi_kappa}) 
give exact solutions:
\bea
   & & \Phi = C, 
   \label{Phi-sol-5} \\
   & & \varphi_\chi = \varphi_+ C + \varphi_d \tilde d 
       \; = \; {3 \over k^2 - 3} {\delta_v \over 1 - \cos{\eta}}, 
   \label{varphi_chi-sol-5} \\
   & & \varphi_\delta = C + (k^2 / 9) (1 - \cos{\eta}) \varphi_\chi,
   \label{varphi_delta-sol-5}
\eea
and similarly for $\varphi_\kappa$;
$\tilde d \equiv d/c_m^2$ is dimensionless.
{}For $\eta \ll 1$ we have $\varphi_\chi = {3 \over 5} C$ and
$\varphi_\delta = \varphi_\kappa = C$ for the $C$-modes.
{}From eq. (\ref{Phi-varphi_chi-2}) we have
 $\dot \Phi = 0$ exactly for a pressureless fluid 
considering general $K$ and $\Lambda$.
$\Phi$ has only the $C$-mode (it is identified as $C$), and no $d$-mode.
Evolutions of $\varphi_\chi$ and $\Phi$ are
shown in Fig. \ref{Fig-pressureless-pert}.
Notice that both $\varphi_+$ and $\varphi_d$ diverge as the model 
approaches the big crunch singularity.

Let us consider a scenario where the big crunch is succeeded to
an expanding phase: thus we have two phases  
$\eta \le \eta_1$ (phase $I$) and $\eta \ge \eta_1$ (phase $II$)
with $\eta_1 = 0$. 
{}For $\varphi_\chi$ we can take two out of three forms of solutions 
($\varphi_+$, $\varphi_-$, $\varphi_d$) as the general solutions
in either phase.
Although $\dot a$ could be discontinuous at the transition reaching
the singularity let us try matching $\varphi_\chi$ and $\varphi_\delta$ 
directly at $\eta_1$; afterall we assume a nonsingular bounce
near $\eta_1$, see later.
Using eqs. (\ref{varphi_chi-sol-5},\ref{varphi_delta-sol-5},\ref{asymptotes})
we can show
\bea
   & & C_{II} = C_{I}, \quad
       \tilde d_{II} = \tilde d_I - 18 \pi C_I.
   \label{C-4}
\eea
This implies that the value of $\Phi$ variable is
conserved even through the bounce.
In terms of the general solutions we notice the following.
Using eq. (\ref{varphi_-}) we can decompose $\varphi_+$ in 
eq. (\ref{varphi_chi-sol-5}) to $\varphi_-$ and $\varphi_d$.
Near bounce, although $\varphi_-^I \simeq {3 \over 5}$ is negligible
compared with $\varphi_+^I \simeq 18 \pi (8/|\eta|^5)$ and
$\varphi_d^I \simeq - 8/|\eta|^5$, we can show that it is this constant 
mode of $\varphi_-^I$ which feeds the $C$-mode after the bounce.
Thus, in the collapsing phase it is appropriate to write 
eq. (\ref{varphi_chi-sol-5}) as
\bea
   & & \varphi_\chi = \varphi_- C + \varphi_d (\tilde d - 18 \pi C).
\eea
Therefore, if such a bounce is allowed, 
we have shown that $\varphi_-^I$ feeds the growing mode 
$\varphi_+^{II}$ in the expanding phase.
The apparent growing (diverging) solutions $\varphi_+^I$
or $\varphi_d^I$ only feed the $\varphi_d^{II}$ or $\varphi_-^{II}$
which are the decaying mode in expanding phase.
The time-scale of a cycle is encoded in $c_m$ of eq. (\ref{BG-sol})
which can affect only the decaying solution in the expanding phase.
The value of $\Phi$, which is $C$, is not affected by the
different duration of each cycle.

Notice, however, that if we strictly consider the singular and cuspy 
bounce at $\eta_1$ implied by eq. (\ref{BG-sol}) we have $\dot a$ 
discontinuous, which forbids us from relying on the matching conditions.
We have assumed such a singular bounce can be regarded as a limiting 
case of a smooth and non-singular bounce; a concrete example
will be considered in the next section. 
We note that the curvature term has negligible role near the big crunch/bang.
We also have assumed the linearity of the fluctuations involved.

\section{Through the bounce}
                                              \label{sec:Bounce}

Assume two expanding phases $I$ and $II$ with equation of states 
$w_I$ and $w_{II}$.
In near flat situation we have
\bea
   & & a (t) = a_{0i} (t - t_{i})^{2 \over 3 (1 + w_{i})},
\eea
where $i = I, II$.
The coefficients should be determined by matching $a$ and $\dot a$ 
at the transitions $t_1$.
In phases $I$ and $II$ the large-scale solutions in 
eqs. (\ref{Phi-sol},\ref{varphi_chi-sol},\ref{varphi_delta}) give
for $-1 < w_i < 1$:
\bea
   & & \Phi = C + k^2 {4 \over 9} {w_i \over 1 - w_i^2} {1 \over a^3 H} d, 
   \label{Phi-sol-3} \\
   & & \varphi_\chi = {3 + 3w_i \over 5 + 3w_i} C + {H \over a} d,
   \label{varphi_chi-sol-3} \\
   & & \varphi_\delta = C + k^2 {2 \over 9} {1 \over 1 - w_i} {1 \over a^3 H} d,
   \label{varphi_delta-sol-3} 
\eea 
where $C$ and $d$ in the phase $I$ should be regarded as $C_I$ and $d_I$,
and similarly for the phase $II$.
{}For $w_i = 1$, $d$-mode of $\Phi$ (and part of $\varphi_\delta$) contains
$\ln{(t-t_i)}$ term instead of $(1 - w_i)^{-1}$.
Equations (\ref{Phi},\ref{varphi_kappa}) give $\varphi_v$ and $\varphi_\kappa$.
As long as we have $a$ and $\dot a$ continuous through a transition
from phase $I$ to phase $II$ at $t_1$ we can use our joining variables
in eq. (\ref{matching-variables}).
Examples are the radiation-matter transition and
the inflation-radiation transition.
Using eqs. (\ref{varphi_chi-sol-3},\ref{varphi_delta-sol-3}), 
to the leading order in the large-scale expansion we have
\bea
   & & C_{II} = C_{I}, 
   \nonumber \\
   & & d_{II} = d_I + {6 (w_I - w_{II}) \over (5 + 3 w_I)(5 + 3 w_{II})}
       {a_1 \over H_1} C_{I}.
   \label{C-2}
\eea
This is consistent with the result derived in eq. (17) of 
\cite{Hwang-Vishniac-1991}.
Similar results hold for two contracting phases as well.

In the case of transition from the contracting to expanding phases,
however, $\dot a$ can be discontinuous at the transition.
In order to handle the case properly, we need an intermediate 
bouncing phase $B$ which smoothly connects the two phases $I$ and $II$.
We consider the collapsing ($I$) and expanding ($II$) phases 
smoothly connected by a nonsingular bouncing phase ($B$).
Assuming the curvature is not important in phases $I$ and $II$
near the bounce, and assuming $w_I$ and $w_{II}$ for the two phases we have
\bea
   & & a_{I} (t) = a_{I0} [ - (t - t_{I}) ]^{2 \over 3 (1 + w_{I})},
   \label{BG-a-I} \\
   & & a_{II} (t) = a_{II0} (t - t_{II})^{2 \over 3 (1 + w_{II})}.
   \label{BG-a-II} 
\eea
The coefficients should be determined by matching $a$ and $\dot a$ 
at the transitions $t_1$ and $t_2$.

In the expanding phase $II$, the $d$-modes of $\Phi$, $\varphi_\chi$,
$\varphi_\delta$, $\varphi_v$ and $\varphi_\kappa$ in 
eqs. (\ref{Phi-sol-3}-\ref{varphi_delta-sol-3}) decay away whereas $C$-modes 
remain constant and have the roles of the relatively growing modes.
Whereas, as $t \rightarrow t_I$ in the contracting phase
although the $C$-modes remain constant,
the $d$-modes diverge (for $-1< w_I < 1$) as
\bea
   & & \Phi \propto \varphi_v \propto \varphi_{\delta} \propto \varphi_\kappa 
       \propto {1 \over a^3 H} 
       \propto |t-t_I|^{-{1-w_I \over 1+w_I}},
   \nonumber \\
   & & \varphi_{\chi} \propto {H \over a} 
       \propto |t-t_I|^{-{5+3w_I \over 3+3w_I}}.
   \label{LS-sol-w}
\eea
{}For $w_I = 0$, the $d$-mode of $\Phi$ vanishes exactly, and
the $d$-mode of $\varphi_v$ vanishes in near flat situation.
{}For $w_I = 1$, the $d$-modes of $\Phi$, $\varphi_v$, $\varphi_\delta$ 
and $\varphi_\kappa$ show $\ln{|t-t_I|}$ divergence, instead.
A complete set of solutions in several different gauge conditions is
presented in Tables 2-5 of \cite{Hwang-1993-IF}; although the solutions
were derived in the expanding phase, the same solutions remain valid
in the collapsing phase with the time replaced by its absolute value.

\begin{figure}[ht]
   \centering
   \leavevmode
   \epsfysize=8cm
   \epsfbox{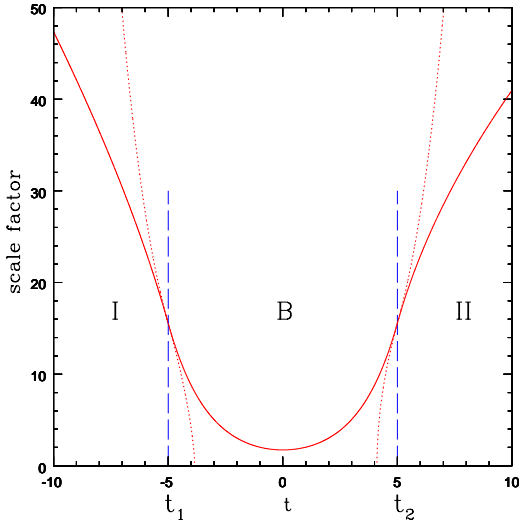}\\
   \caption[fig]
   {\label{Fig-BG}
    Evolution of the scale factor (line).
    The dotted lines indicate the extensions of the collapsing,
    bouncing and expanding phases without matchings. 
    The bouncing phase lasts for $|t_B| <5$ with $\Lambda = 1$ and $K = 1$.
    As an example, we take $w_{I} = 0$ and $w_{II} = {1 \over 3}$.
    }
\end{figure}

A simple example of the bounce is the case with
$K > 0$ and a positive $\Lambda$ \cite{deSitter-1917}
\bea
   & & a_B (t) = \sqrt{3 K / \Lambda} 
       \cosh{( \sqrt{\Lambda / 3} t )}.
   \label{BG-a-B} 
\eea
Evolution of the scale factor is plotted in Fig. \ref{Fig-BG}.
Either for the vanishing fluid with pure $\Lambda$, 
or for a $\Lambda$-type fluid, we have $\mu + p = 0$.
If we have $\mu + p = 0$ strictly the basic set of perturbation equations
becomes trivial and we cannot determine the perturbations properly,
i.e., we do not have meaningful perturbations.

In the following we consider the perturbation evolutions
in three other examples of the bouncing phase.
The first two models rely on the fluid/field which give effectively
$w<-{1 \over 3}$ equation of state during the bounce.
To have the bounce in such models the positive curvature should have
significant role during the bounce, thus not suitable for the
bouncing model assumed in \S \ref{sec:Pressureless}.
In addition, as the background curvature becomes important,
all the perturbation scales go through the small-scale regime
where the pressure gradient term becomes important.
The third model relies on a presence of an exotic matter which
gives a negative contribution to the total energy density.
In this case we have a bounce without resorting to the 
positive curvature, thus the scales remain in the large-scale
during the bounce and the model suits the requirement of the
bounce assumed in \S \ref{sec:Pressureless}.

\subsection{Bounce with a $w = - {2 \over 3}$ fluid}
                                              \label{sec:Fluid-bounce}

\begin{figure}[ht]
   \centering
   \leavevmode
   \epsfysize=8cm
   \epsfbox{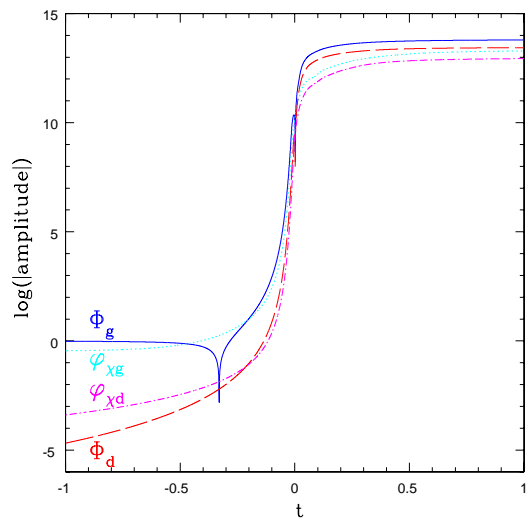} \\
   \caption[fig]
   {\label{Fig-fluid-n=10}
    Evolutions of $\Phi$ for the $C$-mode (blue, line) and 
    $d$-mode (red, long-dashed line)
    initial conditions,
    and $\varphi_\chi$ for the $C$-mode (cyan, dotted line) 
    and $d$-mode (magenta, dot-short-dashed line)
    initial conditions.
    We take $n=10$.
    The pressure terms become important in $|t|<.4$ for $\varphi_\chi$
    and in $|t|<.2$ for $\Phi$.
    Notice that the $C$-mode of $\Phi$ changes sign twice,
    whereas the $d$-mode changes once.
    The sign changes in the expanding phase occur at the same time.
    }
\end{figure}

{}For an ideal fluid with $w = {\rm constant}$, $K > 0$ and $\Lambda = 0$, 
from the Friedmann equation we have
$H = 0$ at $a (t_*) = (2 c_0/K)^{1/(1 + 3 w)}$ where
$c_0 \equiv (4 \pi G/3) \mu a^{3(1 + w)}$.
We can show that
$a (t_*)$ is a maximum for $w > - {1 \over 3}$, and
is a minimum for $w < - {1 \over 3}$.
As a simple example which gives a bounce we consider $w = -{2 \over 3}$ case,
\cite{Zeldovich-etal-1974}.
Although it is uncertain whether it is appropriate to consider an ideal fluid
for $w < 0$ case, we will take the ideal fluid assumption,
see a cautionary remark in \S VII of \cite{Bardeen-1980}.
We will find a fundamentally different result in a more realistic 
(in the sense that we have the concrete action and equation of motion)
case based on a scalar field, see \S \ref{sec:MSF-bounce}.
Equation (\ref{BG-eqs}) gives an exact solution
\bea
   & & a = (c_0/2) (t^2 + K/c_0^2).
\eea
We have $c_s^2 = - {2 \over 3}$ and
\bea
   & & {\ddot x \over x} = {c_0^2 \over 2 a^2} {3 t^4 + (K/c_0^2)^2
        \over t^2}, \quad
       {\ddot y \over y} = {c_0^2 \over 2 a^2} ( t^2 - 3 K/c_0^2 ).
\eea
The pressure terms become important compared with the gravity near the bounce.
Thus, in order to follow the evolutions we need to handle perturbations
based on the full equations.
Equation (\ref{varphi_chi-eq}) gives
\bea
   & & \ddot \varphi_\chi + {4 t \over t^2 + K/c_0^2} \dot \varphi_\chi
       - {8 \over 3 c_0^2} {k^2 - 3 K \over ( t^2 + K/c_0^2 )^2} \varphi_\chi
       = 0.
\eea
Ignoring the $k^2$ term the exact solutions in 
eqs. (\ref{Phi-sol},\ref{varphi_chi-sol}) can be integrated,
and for $\varphi_\chi$ we have
\bea
   & & \varphi_\chi
        = C {1 \over 3} {t^4 + 6 (K/c_0^2) t^2
        - 3 (K/c_0^2)^2 \over (t^2 + K/c_0^2)^2} 
   \nonumber \\
   & & \qquad
       + d {4t/c_0 \over (t^2 + K/c_0^2)^2}.
\eea
Since the $k^2$ terms become negligible away from the bounce we can
use these solutions as the proper initial conditions for the
$C$- and $d$-modes.
A typical evolution is presented in Fig. \ref{Fig-fluid-n=10}.

As we have $c_s^2 = - {2 \over 3}$ we anticipate an exponential
growth/decay of the perturbation while the pressure gradient
term becomes important.
In Fig. \ref{Fig-fluid-n=10} both the $C$- and $d$-modes in the
contracting phase become the (relatively growing) $C$-mode 
in the expanding phase.
As the large-scale conditions are violated both the $C$- and $d$-modes
will be dominated by the exponentially growing mode in the small-scale
limit, and eventually we cannot trace the $C$- and $d$-modes
in the expanding phase to the ones in the contracting phase.

\subsection{Bounce with a massive scalar field}
                                              \label{sec:MSF-bounce}

\begin{figure}[ht]
   \centering
   \leavevmode
   \epsfysize=8cm
   \epsfbox{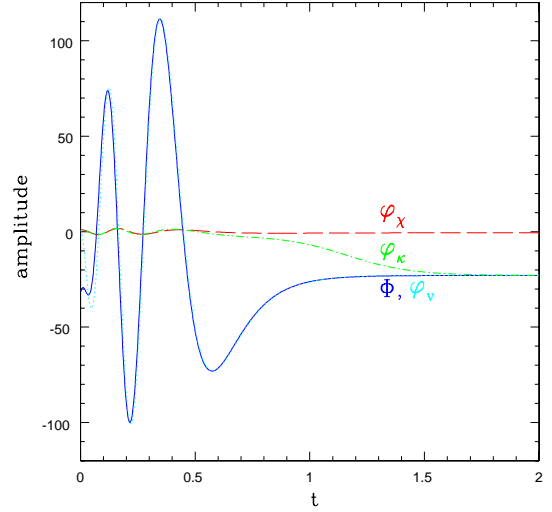}
   \caption[fig]
   {\label{Fig-MSF-n=10}
   Evolutions of $\varphi_\chi$ (red, long-dashed line),
         $\Phi$ (blue, line), $\varphi_v$ (cyan, dotted line), 
         and $\varphi_\kappa$ (green, dot-short-dashed line).
   We omit $\varphi_\delta$ which has a larger amplitude and
   approaches $\Phi$ in the large-scale.
   We take $n = 10$.
   $t = 0$ corresponds to the minimum of the bounce,
   and the pressure dominates over the gravity till $t \sim 1$.
   We can show that $\Phi$, $\varphi_v$, $\varphi_\kappa$
   and $\varphi_\delta$ stay constant in the large-scale,
   whereas $\varphi_\chi$ still adjusting its value even in the large-scale
   as the background equation of state effectively changes from 
   $w_\phi \simeq -1$ while $\phi$ is rolling
   to $w_\phi = 0$ as $\phi$ starts oscillating
   which occurs for $t>11$.
    }
\end{figure}

We consider a bouncing model based on a massive minimally coupled scalar field
with a positive curvature \cite{Barrow-Matzner-1980}.
Results up to eq. (\ref{pressure}) in \S \ref{sec:Review} remain
valid for the field with a prescription mentioned below eq. (\ref{Poisson-eq}). 
The background equations are presented in eq. (\ref{BG-eqs}) and below it
with $\mu_\phi = {1 \over 2} (\dot \phi^2 + m^2 \phi^2)$, etc.
Equation (\ref{varphi_chi-eq}) gives
\bea
   & & \ddot \varphi_\chi + \left( 7 H + 2 {m^2 \phi \over \dot \phi}
       \right) \dot \varphi_\chi
   \nonumber \\
   & & \quad
       + \left( {m^2 \phi^2 \over M_{pl}^2} + 2 H {m^2 \phi \over \dot \phi}
       +  {k^2 - 8 K \over a^2} \right) \varphi_\chi = 0,
   \label{varphi_chi-eq-MSF}
\eea
where $M_{pl}^2 \equiv 1/(8 \pi G)$.
Once we have $\varphi_\chi$, the rest of perturbations
$\Phi$, $\varphi_v$, $\varphi_\delta$ and $\varphi_\kappa$ follow from
eqs. (\ref{Phi},\ref{Phi-varphi_chi-1},\ref{varphi_delta},\ref{varphi_kappa}).
We have
\bea
   & & c_A^2 {k^2 \over a^2} =
       {2 m^2 \phi \over H \dot \phi} {K \over a^2}
       + {k^2 \over a^2},
   \label{p-msf} \\
   & & {\ddot y \over y} = {2 m^2 \phi \over H \dot \phi} {K \over a^2}
       + m^2 - {3 \over 4} {K \over a^2}
       + {7 \dot \phi^2 + 25 m^2 \phi^2 \over 24 M_{pl}^2}
   \nonumber \\
   & & \qquad
       + 2 m^2 {\phi \over \dot \phi} \left( m^2 {\phi \over \dot \phi}
       + 4 H \right).
   \label{g-msf}
\eea
We can show that near the bouncing era the pressure term in eq. (\ref{p-msf})
dominates over the gravity term in eq. (\ref{g-msf});
this is true even for $k^2 = 0$.
Thus, near the bounce the large-scale condition is violated, and
with the positive sign in front of $k^2$ term in eq. (\ref{p-msf})
we can show that perturbations show oscillatory behavior while in the
small scale, see Fig. \ref{Fig-MSF-n=10}.
Although the first term in the RHS of eq. (\ref{p-msf}) diverges
near the bounce, the same term appears in the gravity part in
eq. (\ref{g-msf}) as well.
Due to the positive sign in the second term we expect oscillatory
instability as the term dominates the gravity part.
Near the bounce we have $H \simeq 0$, thus $\mu_\phi \simeq {\rm constant}$,
and
\bea
   & & a \simeq \sqrt{3 M_{pl}^2 K/\mu_\phi} 
       \cosh{\left[ \sqrt{\mu_\phi/(3 M_{pl}^2)} t \right]}.
\eea

In this model, during the bounce all scales reach the small-scale 
where we cannot apply our large-scale solutions.
In Fig. \ref{Fig-MSF-n=10} we used an arbitrary initial
condition at the minimum of the bounce ($t=0$),
and as the scale becomes large-scale we have only the
$C$-mode because the $d$-mode
in expanding phase is decaying (thus transient) and
yields to the relatively growing $C$-mode within a few expansion time scale. 
Thus, as in the previous example based on the $w = - {2 \over 3}$ fluid
the $C$- and $d$-modes during the collapsing phase
are mixed up while in the small-scale, and we cannot trace the
$C$- and $d$-modes in the expanding phase to the ones in the
collapsing phase.
One important difference of the scalar field compared with
the $w = - {2 \over 3}$ fluid is that, while fluctuation of the
fluid shows exponential instability due to the negative
$c_s^2$ term, the field fluctuation shows oscillatory behavior
\cite{Gratton-etal-2001}.
Such a difference comes from the presence of a nonvanishing entropic
perturbation term $e$ in the case of a scalar field,
see below eq. (\ref{Poisson-eq}).

In later expanding phase as $\phi$ starts oscillating near potential
minimum the background model enters an era with effectively
$w_\phi = 0$ (dust) equation of state. 
As $\phi$ starts oscillating we cannot solve 
eq. (\ref{varphi_chi-eq-MSF}) directly.  
Instead, we can analytically handle the situation using proper time averaging
over the coherent oscillations of the background and the
perturbed field \cite{Ratra-1991}.
In \cite{Hwang-1997-axion} it was shown that while the background 
enters the dust era, the perturbations also behave like a cold dark 
matter even in the large-scale limit.

\subsection{Bounce model with $\mu = \mu_m - \mu_X$}
                                         \label{sec:Desparate-bounce}

The positively curved FLRW world model with only the radiation and matter 
does not allow bouncing after the big crunch.
If the physical state near the big crunch allows a presence of additional
matter $X$ with its effective energy density behaving as
$- \mu_X (t) =  - \mu_{X0} a^{-3 (1 + w_X)}$ and $w_X > {1 \over 3}$,
we could have a smooth and non-singular bouncing phase; 
this is a generalized case of a `desparate' example 
mentioned in p368 of \cite{Peebles-1993}.
Thus, for the bounce only, we do not even need the positive curvature 
in the background world model.

As a toy model allowing such a smooth and non-singular transition
with the relevant scale satisfying the large-scale condition
we consider a case with the pressureless matter and the exotic matter
with $w_X = {1 \over 3}$. 
Thus, we consider a model with the dust and the radition with a
negative sign in the radiation component.
\bea
   & & H^2 = {8 \pi G \over 3} (\mu_m - \mu_X) - {K \over a^2}, \quad
   \label{BG-eqs-exotic}
\eea
where $\mu_m \propto a^{-3}$ and $\mu_X \propto a^{-4}$.
Certainly, this is not a realistic model for the bounce because we
need to assume that there is no conventional radiation component present;
to be more realistic, the sum of the radiation and the $X$-matter
should give a net negative contribution.
Later we will show, however, that this toy model captures the basic
physics of more realistic situations.

{}For a positive curvature, $K > 0$, eq. (\ref{BG-eqs-exotic}) gives
an exact solution
\bea
   & & a = {1 \over K} c_m \left[ 1 - \sqrt{1 - 2 (c_X/c_m^2) K}
       \cos{(\sqrt{K}\eta)} \right],
\eea
where $c_X \equiv (4 \pi G/3) \mu_X a^4$; $c_X/c_m^2$ is dimensionless.
We have normalized the time axis so that $a$ has minimum values
at $\eta = 2 \bar n \pi/\sqrt{K}$ with $\bar n$ an integer number.
{}For vanishing $X$-component, $c_X = 0$, we recover the
solution in eq. (\ref{BG-sol}).
With the $X$-component the model shows a cyclic behavior.
The basic picture of the cyclic bounces remains valid in more
realistic situations with $w_X > {1 \over 3}$.

The $K$ term becomes important near $a_{\rm max}$,
and near $a_{\rm min}$ we have $\mu_X \simeq \mu_m$.
The curvature term is negligible near the bounce,
thus allowing existence of the large-scale where we could
ignore the Laplacian term coming from the pressure gradient.
If we do not need a cyclic behavior we can actually take
a flat model with $K = 0$.
Assuming negligible $K$ term we have
\bea
   & & a \simeq {c_m \over 2} \left( \eta^2 + 2 c_X /c_m^2 \right),
\eea
which is an exact solution of eq. (\ref{BG-eqs-exotic}) for $K = 0$.

It is not entirely clear how to handle the perturbation of the
exotic component $X$ which is introduced to have a bounce in the
background without resorting to the positive curvature.
If such an exotic state of matter can be modelled by using a field or
modifying terms in the gravity theory (coming from 
quantum corrections or higher dimensions, for example), 
we need the correct forms of the equation of motion or the modified 
action to handle the perturbations properly.
In the present situation lacking the concrete model for the exotic matter,
we will take a phenomenological approach based on a fluid approximation.

Let us {\it assume}, except for its negative contribution
to the density, the fluid $X$ behaves as an ordinary
ideal fluid with an equation of state $w_X$.
In such a case, near the bounce we need to consider two-component
system with $m$ and $X$.
[The other possibility is to derive solutions in the $m$- and $X$-dominated 
eras separately and to connect by using the matching conditions.  
As the $X$-fluid cannot dominate the total fluid (we have $\mu_m \ge \mu_X$
with the equality holding at the bounce) such a situation is forbidden.]

We use the conventional decomposition of the system into the
adiabatic (curvature) perturbation, characterized by
$\varphi_\chi$ (or $\Phi$), and the relative (often called 
the isocurvature) perturbation defined as 
$S \equiv S_{12} \equiv \delta_1/(1 + w_1) - \delta_2/(1+w_2)$, 
\cite{Kodama-Sasaki-1984}.
The basic equations in the two-component system become, 
see eqs. (23,35,46,57) in \cite{Hwang-Noh-2001-Fluids}
\bea
   & & {\mu + p \over H} \left[ {H^2 \over a (\mu + p)} \left( {a \over H}
       \varphi_\chi \right)^\cdot \right]^\cdot
       + c_s^2 {k^2 \over a^2} \varphi_\chi
       = - 4 \pi G e
   \nonumber \\
   & & \qquad
       = - 4 \pi G { ( \mu_1 + p_1 )
       ( \mu_2 + p_2 ) \over \mu + p } ( c_{(1)}^2 - c_{(2)}^2 ) S,
   \label{varphi_chi-eq-two} \\
   & & \ddot S + H \left ( 2 - 3 c_z^2 \right) \dot S
       + c_z^2 {k^2 \over a^2} S 
   \nonumber \\
   & & \qquad
       = - {k^2 (k^2 - 3K) \over a^4} 
       {c_{(1)}^2 - c_{(2)}^2 \over 4 \pi G (\mu + p)} \varphi_\chi,
   \label{S-eq}
\eea
where $c_{(i)}^2 \equiv \dot p_i/\dot \mu_i$, and
\bea
   & & c_z^2 \equiv {c_{(2)}^2 ( \mu_1 + p_1 )
       + c_{(1)}^2 ( \mu_2 + p_2 ) \over \mu + p}.
\eea
In our case $1 = m$ and $2 = X$, thus, 
$w_1 = c_{(1)}^2 = 0$, $w_2 = c_{(2)}^2 = {1 \over 3}$,
$\mu + p = \mu_m - {4 \over 3} \mu_X$, etc.
In the large-scale limit the curvature mode can be sourced by 
the isocurvature one, see eq. (\ref{varphi_chi-eq-two}),
whereas the isocurvature mode decouples from the curvature mode 
in general \cite{Kodama-Sasaki-1986}, see eq. (\ref{S-eq}).
[The situation is different in the case of multiple number of
scalar field: in such a case, effectively, the RHS of 
eq. (\ref{S-eq}) has $k^2/a^2$ factor instead of $k^4/a^4$,
thus the isocurvature modes are less decoupled from the adiabatic one,
see \cite{Hwang-Noh-2001-Fluids,Hwang-Noh-2000-MSFs}.] 

Let us {\it assume} an adiabatic initial condition, thus setting $S = 0$
at early era in the contracting phase, for simplicity.
More precisely, we are assuming $S \ll \varphi_\chi$ in the initial epoch
which is natural because it means we are assuming no significant fluctuations
in the $X$-component in the early matter dominated era.
Since the curvature mode does not source the isocurvature mode
in the large-scale, the isocurvature mode will remain small.
In such a case the RHS of eq. (\ref{varphi_chi-eq-two}) vanishes, 
and the curvature equations in the single component
situation, eq. (\ref{varphi_chi-eq}), remain valid without any change.
Thus, for scales satisfying the large-scale limit we have
the same solutions in eqs. (\ref{Phi-sol},\ref{varphi_chi-sol}) 
remaining valid.
The solutions become:
\bea
   & & \Phi = C + \tilde d k^2 {64 \over 27} {c_X \over c_m^2}
       \int^\eta \left( \eta^4 + {4 c_X \over 3 c_m^2} \eta^2
       - {4 c_X^2 \over 3 c_m^4} \right)^{-2} \eta^2 d \eta,
   \nonumber \\
   & & \varphi_\chi = 
       \Bigg[ \left( 1 + {c_m^2 \over 2 c_X} \eta^2
       + {5 c_m^4 \over 12 c_X^2} \eta^4
       + {3 c_m^6 \over 40 c_X^3} \eta^6 \right) C
   \nonumber \\
   & & \qquad
       + {c_m^6 \over c_X^3} \eta \tilde d \Bigg] \Bigg/
       \left( 1 + {c_m^2 \over 2 c_X} \eta^2 \right)^{3}.
   \label{sol-mX}
\eea
{}For $\varphi_\chi$ the contribution from the lower bound of 
integration of the $C$-mode is absorbed to the $\tilde d$-mode.
In the matter-dominated eras, $|\eta| \gg 2 c_X/c_m^2$, we have
\bea
   & & \Phi = C - k^2 {64 \over 135} {c_X \over c_m^2} \eta^{-5} \tilde d, 
       \quad
       \varphi_\chi = {3 \over 5} C + 8 \eta^{-5} \tilde d.
   \label{sol-mde}
\eea
Near the big bang/crunch, the solution for $\varphi_\chi$ coincides with 
the one for a pressureless medium considered in eq. (\ref{varphi_chi-sol-5}).
{}For $c_X/c_m^2 \rightarrow 0$, $\Phi$ also coincides with the
one known in the pressureless medium.
Near the bounce, $|\eta| \ll 2 c_X/c_m^2$, we have
\bea
   & & \Phi = C + k^2 {4 \over 9} {c_m^6 \over c_X^3} \eta^3 \tilde d, \quad
       \varphi_\chi = C + {c_m^6 \over c_X^3} \eta \tilde d,
   \label{sol-Xde}
\eea
which are regular and finite.

Since the present bounce model allows the scales to stay in the large-scale 
limit during the transition, it can be considered as a concrete model of
the smooth and non-singular bounce assumed in \S \ref{sec:Pressureless}.
Indeed, the curvature term is negligible near the bounce
as it was the case near the big crunch/bang in \S \ref{sec:Pressureless}.
In the $c_X/c_m^2 \rightarrow 0$ limit, 
eq. (\ref{sol-mX}) reduces to eq. (\ref{sol-mde}) 
which also coincides with the known solution
considered in \S \ref{sec:Pressureless}.

Apparently, we can also make a more realistic model with
the radiation, matter and $X$ where $w_X > {1 \over 3}$.
We note that eqs. (\ref{varphi_chi-eq-two},\ref{S-eq}) remain valid
for any two-component system of matter perturbations.
Even in such a case the $X$ fluid can cause a smooth and non-singular 
bounce and the curvature term has negligible role near the bounce.
Thus, essentially the same conclusions 
(e.g., the $C$-mode feeding the $C$-mode) remain valid.
We have considered a simple toy model only because it allows
analytic handling of the background and the perturbations,
thus showing the situation explicitly.

\section{Discussions}
                                              \label{sec:Discussions}

In \S \ref{sec:Pressureless} we have shown that the perturbation
in a positively curved FLRW model filled with a pressureless matter 
is described by the conservation of $\Phi$.
Assuming a transition of big crunch followed by a big bang
in such a model, by using the known matching conditions 
we have shown that $\Phi$ maintains the same value even after the transition.
Using the matching conditions we also have shown that the diverging
solution in the contracting phase is matched to the decaying
solution in the subsequent expanding phase, whereas the other
solution which stays constant during the contracting phase is
matched into the same constant solution in the expanding phase.
That constant mode is characterized by $\Phi$; the other solution
of $\Phi$ which decays in expanding phase 
is higher-order in $k^2$ compared with the one of
$\varphi_\chi$ and vanishes for the vanishing background pressure.

In order to confirm these results based on the matching
at singular bounce, in \S \ref{sec:Bounce} we have considered
three different non-singular and smooth bounce models.
{}For the bounce models based on the fluid (\S \ref{sec:Fluid-bounce})
and the massive scalar field (\S \ref{sec:MSF-bounce}) in the
positively curved background, the role of background curvature is 
important to make the bounce.
In such cases, all the perturbation scales
come inside the sound-horizon near bounce, 
and the large-scale conditions are violated.
As the pressure gradient terms become important, perturbations in the fluid 
model show exponential instability, whereas the ones in the massive field model
show oscillatory behaviors.
{}For both situations the two independent perturbation modes
in the large-scale limit during the collapsing phase got mixed up
with the two independent modes in the small-scale during the bouncing phase.
Thus, we cannot trace the two independent solutions ($C$- and
$d$-modes) in the expanding phase to the ones in the contracting phase.

In \S \ref{sec:Desparate-bounce} we considered a bouncing model
based on an exotic matter with negative contribution to the total 
energy density.
In such a case the positive curvature is not important during the bounce.
Thus, we could have the relevant scales remaining in the
large-scale limit, and could apply the general large-scale solutions. 
In this case, however, we have to handle the perturbation of the
exotic matter in addition to the ordinary one simultaneously.
By considering the adiabatic initial condition we have shown that
the same curvature perturbation equation known in the single-component
situation remains valid, thus the known large-scale solutions
are valid as well throughout the bounce. 
Therefore, this third-type of bounce model can be regarded as an example of
the smooth and non-singular bouncing assumed in \S \ref{sec:Pressureless}.
As an analytically manageable concrete example, 
we have considered a simple toy model with a dust and an exotic matter 
with the radiation-like equation of state.
Even in more general situations of the equation of states
before and after the bounce, similar analyses can be made which show 
that the $C$-mode in the expanding phase is affected 
only by the $C$-mode in the contracting phase, thus the growing mode in 
the contracting phase decays away as the world model enters expanding phase.

Our analyses are based on two assumptions:
(i)  the contracting phase is converted into the expanding one
     by a smooth and non-singular bounce, and
(ii) the linear perturbation theory holds during the evolution.
The large-scale evolution can be characterized by the conservation of $\Phi$.
We have shown that the $C$-mode of $\Phi$, which is the proper
growing mode in expanding phase, is simply conserved during the
evolution and through bounces.
The results are true as long as the two assumptions made above are valid,
and in addition, if the large-scale condition is met during the transition
as considered in \S \ref{sec:Desparate-bounce}.

In \S \ref{sec:Solutions} we showed that the three dimensionless measures, 
the intrinsic curvatures ($\varphi$
and $C^{(t)}_{\alpha\beta}$), the tracefree part of the
extrinsic curvature ($\hat \sigma/H$), and the Weyl curvature ($E/R$),
{\it diverge} at singularity for $-1<w<1$.
Thus, for $-1<w<1$, the spacetime perturbations become singular as
the background approaches the singularity.  
An ambiguity remains for $w=1$ case because although $\varphi$ and
$C^{(t)}_{\alpha\beta}$ diverge logarithmically, $\hat \sigma/H$
and $E/R$ remain finite.
These apply to all perturbation types and for all
gauge conditions we have considered. 
Behaviors of the other
variables (the perturbed lapse function $\alpha$, a dimensionless
measure of the perturbed expansion $\kappa/H$, relative density
perturbation $\delta$, etc.) depend more strongly on the gauge
conditions, see Tables 2 and 4 of \cite{Hwang-1993-IF}.
Thus, these variables apparently have less physical significance to
characterize the spacetime fluctuations compared with the other
three measures whose behaviors are gauge independent at least in
the pool of gauge conditions we have investigated.
Do above results imply diverging spacetime fluctuations
for $-1<w<1$, and regular ones for $w=1$?
In Table 4 of \cite{Hwang-1993-IF} we find that in no gauge
condition we have {\it all} the perturbations remain finite for $-1<w \le 1$.

The authors of \cite{Peter-Pinto-Neto-2001} argued that as the model goes
through a singular bounce the perturbation becomes nonlinear.
We have shown that {\it if} the fluctuations survive the bounce as
linear ones, the diverging mode in the contracting phase 
should be matched to the decaying one in the expanding phase.
Lyth in \cite{Lyth-2001-2} made the following simple and powerful argument.
As we have under the gauge transformation ($\tilde x^a
= x^a + \xi^a$ with $\xi^t \equiv a \xi^\eta$)
\bea
   & & \tilde \varphi = \varphi - H \xi^t, \quad
       \tilde \delta = \delta + 3 H (1 + w) \xi^t,
   \label{Lyth-conditions}
\eea
if $\varphi$ diverges while $\delta$ remains finite, or vice
versa, in any single gauge condition 
[this is the case for $-1<w \le 1$, 
see eqs. (\ref{Poisson-eq},\ref{LS-sol-w}) for $\varphi_v$ and $\delta_v$]
no temporal gauge-transformation $\xi^t$ can be found which
makes both $\varphi$ and $\delta$ finite. 
Therefore, for $-1<w \le 1$ we encounter the $d$-mode perturbations 
of Friedmann world model becoming {\it singular} near big-crunch 
in one form or another in {\it all} gauge conditions.

We note that $\Phi$, which becomes $\varphi_v$ for $K = 0$,
simply stays constant in a pressureless medium, thus
its magnitude {\it cannot} characterize the breakdown of 
linearity of the perturbation.
As we have from eqs. (\ref{Phi},\ref{Phi-sol-5},\ref{varphi_chi-sol-5})
$\varphi_v = C + (1/3) (1 - \cos{\eta}) \varphi_\chi$ where we set $K = 1$,
$\varphi_v$ itself could diverge near singularity.
{}From eq. (\ref{varphi_kappa}), near big crunch in the pressureless 
medium we have $\varphi_\kappa \simeq \varphi_\delta$ where $\varphi_\delta$,
given in eq. (\ref{varphi_delta-sol-5}), has diverging part.
Thus, near big crunch the diverging modes behave as
\bea
   & & \varphi_v \propto \varphi_\delta \propto \varphi_\kappa
       \propto |\eta|^{-3}, \quad
       \varphi_\chi \propto |\eta|^{-5},
   \label{diverging-sol}
\eea
whereas $\Phi$ has no diverging mode in the pressureless case.
{}For the situation with general $w$, see eq. (\ref{LS-sol-w}).
Bardeen has argued that the behavior of
$\varphi_\chi$ ``overstates the physical strength of the singularity'',
see below eq. (5.12) in \cite{Bardeen-1980}. 

At the singular big crunch, we certainly have $d$-modes of many perturbation
variables unambiguously becoming singular for $-1<w \le 1$,
see Tables 2-4 in \cite{Hwang-1993-IF}.
Do large amplitudes of perturbations imply breakdown of linear theory?
Due to the gauge dependence of relativistic perturbation, large
(larger than unity, say) amplitudes of some dimensionless measures of 
gauge-invariant
perturbation variables do not guarantee the breakdown of the linear theory.
However, what Lyth \cite{Lyth-2001-2} has shown is that in the
collapsing phase we could encounter situations where
the amplitudes of perturbation variables become large
in one form or the other in {\it all} gauge conditions.
Lyth has argued this as the violation of the {\it necessary
condition} of the linear perturbation theory.

In our models avoiding the singularity by a 
smooth and nonsingular bounce it is likely that certain scales
can safely go through the bounce retaining their linear nature.
As we have assumed the linearity of perturbations, our analyses 
and results are applicable to such scales only.
In \S \ref{sec:Pressureless} and \ref{sec:Desparate-bounce} 
we have shown that the diverging solutions in the contracting phase
in eq. (\ref{diverging-sol}) affect only the decaying, thus 
transient, solutions in the subsequent expanding phase.
In such a scenario, however, one could anticipate large (compared
with the $C$-mode) amount
of the decaying ($d$) mode present in the early big-bang phase for a while
which is the remnant from the preceding phase before the big-bang.

In a recently proposed ekpyrotic scenario 
based on colliding branes \cite{brane-collision} it was argued that the final 
scalar-type perturbation is scale invariant \cite{ekpyrotic-scenario}.
In \cite{ekpyrotic-pert} it was shown that the generated scale-invariant 
spectrum in the zero-shear gauge during the collapsing phase should
be identified as the $d$-mode, thus after the bounce we have a
different power-spectrum \cite{Lyth-2001-1}.
Our results in this paper confirm that while the large-scale 
condition is met during the (smooth and non-singular) transition
the $d$-mode in the contracting phase does
not affect the (properly growing) $C$-mode in the expanding phase.
The background curvature is flat in the ekpyrotic scenario
and the scale remains in the large-scale during the bounce.
However, since the bounce of the ekpyrotic scenario goes through a singularity 
the author of \cite{Lyth-2001-2} has argued that one cannot rely on 
the linear analyses as the model approaches the singularity.
Thus, either the final spectrum is not scale-invariant 
(which is the case if the linear perturbation survives) 
or the issue should be handled in the future string theory context 
with a concrete mechanism for the bounce.

\subsection*{Acknowledgments}

We thank Ruth Durrer, Christopher Gordon, Antony Lewis, Patrick Peter,
Dominik Schwarz and Ewan Stewart for useful discussions.
We wish to thank George Efstathiou for sharing his knowledge on perturbations,
and interests throughout the work, and 
Neil Turok for incisive comments on several aspects of the work and 
informing us the massive scalar field model as a candidate for the bounce. 
We are also informed that C. Gordon and N. Turok are currently analysing
the massive field model more closely.
We thank Professor James Bardeen for critical comments on
\S \ref{sec:Discussions}.
HN was supported by grant No. 2000-0-113-001-3 from the
Basic Research Program of the Korea Science and Engineering Foundation.
JH was supported by Korea Research Foundation grants (KRF-99-015-DP0443,
2000-015-DP0080 and 2001-041-D00269).

\baselineskip 0pt


\begin{thebibliography}{99}
\bibitem{Tolman-1931}
         R.C. Tolman, Phys. Rev. {\bf 38}, 1758 (1931);
         G. Lema\^itre, Ann. Soc. Sci. Bruxelles A {\bf 53}, 51 (1933),
            translated in Gen Rel. Grav. {\bf 29}, 935 (1997);
         M.J. Rees, The Observatory {\bf 89}, 193 (1969);
         R. Dicke and P.J.E. Peebles, in: {\it General relativity},
            eds., S.W. Hawking and W. Israel,
            (Cambridge Univ. Press, Cambridge, 1979), 504p;
         Ya.B. Zel'dovich and I.D. Novikov,
            {\it Relativistic astrophysics, Vol 2, The structure and
            evolution of the universe} (Univ. Chicago Press, Chicago, 1983);
         G.F.R. Ellis, Ann. Rev. Astron. Astrophys. {\bf 22}, 157 (1984);
         J.D. Barrow and F.J. Tipler, {\it The Anthropic cosmological
              principle} (Clarendon Press, Oxford, 1986);
         J.D. Barrow and M.P. D\c{a}browski, Mon. Not. R. Astron. Soc.
              {\bf 275}, 850 (1995).
\bibitem{Tolman-1934}
         R.C. Tolman, {\it Relativity, thermodynamics and cosmology}, 
               (Oxford Univ. Press, London, 1934).
\bibitem{Peebles-1993}
         P.J.E. Peebles, {\it Principles of physical cosmology}
                  (Princeton Univ. Press, Princeton, 1993).
\bibitem{Nariai-1971}
         H. Nariai, Prog. Theor. Phys. {\bf 46}, 433 (1971);
         H. Nariai and K. Tomita, {\it ib id}. {\bf 46}, 776 (1971);
         L. Parker and S.A. Fulling, Phys. Rev. D {\bf 7}, 2357 (1973);
         G. L. Murphy, {\it ib id}. {\bf 8}, 4231 (1973);
         H. Nariai, Prog. Theor. Phys. {\bf 51}, 613 (1974);
         J.D. Bekenstein, Phys. Rev. D {\bf 11}, 2072 (1975);
         V.N. Melnikov and S.V. Orlov, Phys. Lett. A {\bf 70}, 263 (1979);
         M. Visser, Phys. Lett. B {\bf 349}, 443 (1995), (gr-qc/9409043).
\bibitem{Molina-Paris-Visser-1999}
         C. Molina-Par\'is and M. Visser, Phys. Lett. B {\bf 455}, 90 (1999),
            (gr-qc/9810023).
\bibitem{bounce-cyclic}
         J. Khoury, B.A. Ovrut, N. Seiberg, P.J. Steinhardt, and 
            N. Turok, hep-th/0108187;
         P.J. Steinhardt and N. Turok, hep-th/0111098;
         G. Felder, A. Frolov, L. Kofman, and A. Linde, hep-th/0202017.
\bibitem{Bardeen-1988}
         J.M. Bardeen, in: {\it Cosmology and particle physics},
               eds., L. Fang and A. Zee, (Gordon and Breach, London, 1988), 1;
         J. Hwang, Astrophys. J. {\bf 375}, 443 (1991).
\bibitem{HN-2002-CMBR}
         J. Hwang and H. Noh, Phys. Rev. D {\bf 65} 023512 (2002),
            (astro-ph/0102005).
\bibitem{Bardeen-1980}
         J.M. Bardeen, Phys. Rev. D {\bf 22}, 1882 (1980).
\bibitem{Field-Shepley-1968}
         G.B. Field and L.C. Shepley, Astrophys. Space. Sci.
               {\bf 1}, 309 (1968);
         G.B. Field, in {\it Stars and stellar system, Vol IX,
               Galaxies and the universe}, eds., A. Sandage,
               M. Sandage, and J. Kristian, (Univ. of Chicago Press, 1975)
               Chicago), 359;
         G.V. Chibisov and V.F. Mukhanov, Mon. Not. R. Astron. Soc.
               {\bf 200}, 535 (1982).
\bibitem{Hwang-Vishniac-1990}
         J. Hwang and E.T. Vishniac, Astrophys. J. {\bf 353}, 1 (1990).
\bibitem{Hwang-1999}
         J. Hwang, Phys. Rev. D {\bf 60}, 103512 (1999), (astro-ph/9907080).
\bibitem{Hwang-Noh-2001-Fluids}
         J. Hwang and H. Noh, Class. Quant. Grav. in press (2002),
            astro-ph/0103244.
\bibitem{Liddle-Lyth-2000}
         A.R. Liddle and D.H. Lyth,
               {\it Cosmological inflation and large-scale structure},
               (Cambridge Univ. Press, Cambridge, 2000).
\bibitem{Lukash-1980}
         V.N. Lukash, Sov. Phys. JETP Lett. {\bf 31}, 596 (1980);
               Sov. Phys. JETP {\bf 52}, 807 (1980).
\bibitem{Mukhanov-etal}
         V.F. Mukhanov, JETP Lett. {\bf 41}, 493 (1985);
         V.F. Mukhanov, Soviet Phys. JETP {\bf 68}, 1297 (1988);
         V.F. Mukhanov, H.A. Feldman, and R.H. Brandenberger, 
            Phys. Rep. {\bf 215}, 203 (1992).
\bibitem{Lifshitz-1946}
         E.M. Lifshitz, J. Phys. (USSR) {\bf 10}, 116 (1946);
         E.M. Lifshitz and I.M. Khalatnikov, Adv. Phys. {\bf 12}, 185 (1963);
         L.D. Landau and E.M. Lifshitz, {\it The Classical Theory of Fields},
               4th ed. (Pergamon, Oxford, 1975) \S 115.
\bibitem{Lyth-Woszczyna-1995}
         D.H. Lyth and A. Woszczyna, Phys. Rev. D {\bf 52}, 3338 (1995),
              (astro-ph/9501044).
\bibitem{Hwang-Vishniac-1991}
         J. Hwang and E.T. Vishniac, Astrophys. J. {\bf 382}, 363 (1991).
\bibitem{joining-applications}
         N. Deruelle and V.F. Mukhanov, Phys. Rev. D {\bf 52}, 5549 (1995),
            (gr-qc/9503050);
         J. Martin and D.J. Schwarz, Phys. Rev. D {\bf 57}, 3302 (1998),
            (gr-qc/9704049).
\bibitem{Harrison-1991}
         E. Harrison, Astrophys. J. {\bf 383}, 60 (1991).
\bibitem{Lightman-Press-1989}
         A.P. Lightman, and W.H. Press, Astrophys. J. {\bf 337}, 598 (1989).
\bibitem{Harrison-1970}
         E.R. Harrison, Phys. Rev. D {\bf 1}, 2726 (1970);
         D.H. Lyth, {\it ib id.} {\bf 31}, 1792 (1985);
         D.H. Lyth, and M. Mukherjee, {\it ib id.} {\bf 38}, 485 (1988);
         J. Hwang, Astrophys. J. {\bf 380}, 307 (1991);
         J. Hwang, and J. J. Hyun, {\it ib id.} {\bf 420}, 512 (1994).
\bibitem{Starobinsky-etal-2001}
         A.A. Starobinsky, S. Tsujikawa, and J. Yokoyama,
              Nucl. Phys. B {\bf 610} 383 (2001), (astro-ph/0107555).
\bibitem{Hwang-1993-IF}
         J. Hwang, Astrophys. J. {\bf 415}, 486 (1993).
\bibitem{covariant}
         J. Ehlers, Gen. Rel. Grav. {\bf 25}, 1225 (1993),
            translated from German originally published in
            Akad. Wiss. Lit. Mainz, Abhandl. Math.-Nat. Kl.
            {\bf 11} 792 (1961);
         G.F.R. Ellis, in {\it General relativity and cosmology,
                Proceedings of the international summer school of physics
                Enrico Fermi course 47},
                edited by R. K. Sachs (Academic Press, New York, 1971), 104;
         G.F.R. Ellis, in {\it Cargese Lectures in Physics},
                edited by E. Schatzmann (Gorden and Breach, New York, 1973), 1.
\bibitem{Durrer-1994}
         R. Durrer, Fund. Cosmic Phys. {\bf 15}, 209 (1994),
            (astro-ph/9311041).
\bibitem{Brustein-etal-1995}
         R. Brustein, M. Gasperini, M. Giovannini, V. Mukhanov,
            and G. Veneziano, Phys. Rev. D {\bf 51}, 6744 (1995),
            (hep-th/9501066).
\bibitem{Friedmann-1922}
         A. Friedmann, Zeitschrift f\"ur Physik {\bf 10}, 377 (1922),
            translated in: {\it Cosmological constants, papers in 
            modern cosmology}, eds., J. Bernstein and F. Feinberg,
            (Columbia Univ. Press, New York, 1986), 49p.
\bibitem{Jeans-1902}
         J. Jeans, Phil. Trans. Roy. Soc. {\bf 199A}, 49 (1902);
         W.B. Bonnor, Mon. Not. R. Astron. Soc. {\bf 107}, 104 (1957);
         E.R. Harrison, Rev. Mod. Phys. {\bf 39}, 862 (1967).
\bibitem{Nariai-1969}
         H. Nariai, Prog. Theor. Phys. {\bf 41}, 686 (1969);
         K. Sakai, {\it ib id}. {\bf 41}, 1461 (1969).
\bibitem{Weinberg-1972}
         \S 15.9 in S. Weinberg, {\it Gravitation and cosmology} 
                    (Wiley, New York, 1972);
         \S 11 in P.J.E. Peebles,  
                  {\it The large-scale structure of the universe} 
                  (Princeton Univ. Press, Princeton, 1980).
\bibitem{Groth-Peebles-1975}
         E.J. Groth and P.J.E. Peebles, Astron. Astrophys.
               {\bf 41}, 143 (1975).
\bibitem{deSitter-1917}
         W. de Sitter, Mon. Not. R. Astron. Soc. {\bf LXXVIII}, 3 (1917).
\bibitem{Zeldovich-etal-1974}
         Ya.B. Zel'dovich, L.B. Okun, and I.Yu. Kabzarev,
                Sov. Phys. JETP {\bf 40}, 1 (1975).
\bibitem{Barrow-Matzner-1980}
         J.D. Barrow and R.A. Matzner, Phys. Rev. D {\bf 21}, 336 (1980). 
\bibitem{Gratton-etal-2001}
         S. Gratton, A. Lewis, and N. Turok, astro-ph/0111012. 
\bibitem{Ratra-1991}
         B. Ratra, Phys. Rev. D {\bf 44}, 352 (1991).
\bibitem{Hwang-1997-axion}
         J. Hwang, Phys. Lett. B {\bf 401}, 241 (1997), (astro-ph/ 9610042).
\bibitem{Kodama-Sasaki-1984}
         H. Kodama and M. Sasaki, Prog. Theor. Phys. Suppl.
            {\bf 78}, 1 (1984).
\bibitem{Kodama-Sasaki-1986}
         H. Kodama and M. Sasaki, Int. J. Mod. Phys. A {\bf 1}, 265 (1986).
\bibitem{Hwang-Noh-2000-MSFs}
         J. Hwang and H. Noh, Phys. Letts. B {\bf 495}, 277 (2000),
            (astro-ph/0009268).
\bibitem{Peter-Pinto-Neto-2001}
         P. Peter and N. Pinto-Neto, Phys. Rev. D {\bf 65}, 023513 (2002),
            (gr-qc/0109038).
\bibitem{Lyth-2001-2}
         D.H. Lyth, hep-ph/0110007.
\bibitem{brane-collision}
         C. Park and S-J Sin, Phys. Lett. B {\bf 485}, 239 (2000),
            (hep-th/0005013);
         J. Khoury, B.A. Ovrut, P.J. Steinhardt, and N. Turok, 
            Phys. Rev. D {\bf 64} 123522 (2001), (hep-th/0103239).
\bibitem{ekpyrotic-scenario}
         J. Khoury, B.A. Ovrut, P.J. Steinhardt, and N. Turok, 
            hep-th/0109050;
         R. Durrer, hep-th/0112026.
\bibitem{ekpyrotic-pert}
         R. Brandenberger and F. Finelli, JHEP {\bf 0111}, 056 (2001),
            (hep-th/0109004);
         J. Hwang, Phys. Rev. D in press (2002), (astro-ph/0109045);
         S. Tsujikawa, gr-qc/0110124;
         F. Finelli and R. Brandenberger, hep-th/0112249;
         J. Martin, P. Peter, N. Pinto-Neto, D.J. Schwarz, hep-th/0112128.
\bibitem{Lyth-2001-1}
         D.H. Lyth, Phys. Lett. B {\bf 524}, 1 (2002), (hep-ph/0106153).
\end{thebibliography}
\end{document}